\documentclass[twocolumn,showpacs,preprintnumbers,amsmath,amssymb]{revtex4}

\usepackage{graphicx}
\usepackage{dcolumn}
\usepackage{bm}
\usepackage{amsmath,amssymb,ams,bbm,epsfig}

\begin{document}

\title{Atomic current across an optical lattice}

\author{Alexey V. Ponomarev$^{1,2}$}
\author{Javier Madro\~nero$^{1,3}$}
\author{Andrey R. Kolovsky$^{1,2}$} 
\author{Andreas Buchleitner$^{1}$}
\affiliation{$^{1}$Max-Planck-Institut f\"ur Physik komplexer Systeme, N\"othnitzer Str. 38, D-01187 Dresden \\
$^{2}$Kirensky Institute of Physics, Ru-660036 Krasnoyarsk \\
$^{3}$Physik Department, Technische Universit\"at M\"unchen, James-Franck-Stra{\ss}e, D-85747 Garching}

\date{\today}

\begin{abstract}
We devise a microscopic model for the emergence of a collision-induced, fermionic atomic
current across a tilted optical lattice. Tuning the -- experimentally controllable -- parameters
of the microscopic dynamics allows to switch from Ohmic to
negative differential conductance.
\end{abstract}
\maketitle

Ultracold atoms in optical potentials provide a versatile tool for
the experimental study of many-body dynamics, with an unprecedented degree of
accuracy
\cite{ott04,paredes04,koehl05}. 
Not only do these systems allow the faithful
experimental realization of fundamental model Hamiltonians of solid state
theory, and, thus, lend themselves as efficient analog quantum
simulators \cite{jaksch05}. Beyond that, they bridge the gap between the
single particle and
the thermodynamic limit: if not now, so certainly in the near future
experimentalists will be able to load a precisely controlled number of
particles into engineered potentials, with particle numbers tunable from one
to intermediate or large values \cite{chuu05}. Hence, by continuously
varying the system size, we will be able to control the 
emergence of macroscopic,
thermodynamic observables from perfectly deterministic Hamiltonian dynamics,
in experiment and theory. This holds the potential of a much deeper and
quantitative understanding of quantum statistical laws, together with a
variety of possible applications in the context of the ever advancing
miniaturization of modern technology
\cite{fortagh05}. 

In the present contribution, we address
a specific problem of exemplary importance within this context:
the emergence of a macroscopic current across a periodic potential
under static forcing. It is well known that electrons in a perfectly periodic
lattice exhibit coherent Bloch oscillations, under static forcing, and that no
net transport, i.e., no current across such lattice occurs in the absence
of any (incoherent or dephasing) relaxation process \cite{Anderson92}. Only
the latter will induce a net drift velocity of the electrons, giving rise to a
measurable 
current.
While relaxation processes due to phonon scattering or
impurities are abundant yet largely uncontrolled in solids, a clean and
perfectly controllable realization of this fundamental transport problem is feasible in a quantum
optical setting, where we can engineer
the environment:
By loading spin polarized fermions together with ultracold bosons
into a one dimensional optical
lattice, we establish a strictly analogous, idealized scenario. The
bosons act as a bath for the fermions, with relaxation induced by
fermion-boson collisions. 

Our results are obtained from a microscopic
model for very moderate
system sizes -- typically less than ten bosons interacting with a single
fermion, on approx. ten lattice sites. Yet, due to exponential increase of the
many particle system's 
Hilbert space dimension with boson number $N$ and lattice size $L$, this is
sufficient to enter the regime of asymptotic convergence into the thermodynamic
limit \cite{akabepl04}.
Thus, we describe how a macroscopic current emerges from perfectly Hamiltonian
microscopic dynamics, with all ingredients under essentially perfect
experimental control.

Our model Hamiltonian reads 
\begin{equation}
\label{1}
H=H_{\rm F}+H_{\rm B}+H_{\rm int} \;,
\end{equation}
and decomposes into the (single particle) fermionic part 
\begin{equation}
\label{2}
H_{\rm F}=-\frac{J_F}{2}\left(\sum_l |l+1\rangle\langle l| +h.c.\right)
  +Fd\sum_l |l\rangle l\langle l| \;,
\end{equation}
the (many particle) bosonic part 
\begin{equation}
\label{3}
H_{\rm B}=-\frac{J_B}{2}\left(\sum_l \hat{a}^\dag_{l+1}\hat{a}_l +h.c.\right)
  +\frac{W_B}{2}\sum_l \hat{n}_l(\hat{n_l}-1) \;,
\end{equation}
and a term which mediates the collisional interaction
  between fermions and bosons,
\begin{equation}
\label{4}
H_{\rm int}=W_{\rm FB}\sum_l \hat{n}_l|l\rangle\langle l| \;.
\end{equation}

Since we will be interested in the fermionic dynamics, the fermionic tunnelling
coupling $J_F$ in (2) sets the energy scale, and, correspondingly, $\hbar/J_F$
is 
the natural unit of time. Equations (1-4) describe the dynamics of ultracold
atoms in the lowest 
band of a 1D lattice, assuming short range (on-site) interaction, and
tunnelling 
coupling only between adjacent lattice sites $l$.
 Low temperatures as well as moderate field strengths $Fd<J_F$ are required for
consistency with the single band approximation. 
The
single 
particle ansatz for the fermionic part is justified since the Pauli principle
forbids occupation of the same site.
Our model additionally implies that only fermions experience a static forcing
$F$, 
while the bosons see a periodic potential (period $d$) without static forcing.
Such situation can be realized by a suitable
choice of the internal electronic states the fermions and bosons are initially
prepared in \cite{takasu03}.

We impose $J_B = J_F$ for the bosonic component and focus on
filling factors $\bar{n} =N/L \sim 1$ with $N$ bosons distributed over $L$
lattice sites 
(for more than one fermion, $N$ represents the boson number per fermion).
These filling factors interpolate between single-particle dynamics and the
thermodynamic limit, 
for the lattice sizes we consider. The above choice of parameters is in reach
for state of the art 
experiments: $J_B \simeq J_F/2$ has already been realized in experiments
\cite{modugno03} on a
mixture 
of $^{40}$K and $^{87}$Rb, with $J_F \simeq 0.06 E_R$, and
experimental techniques 
for the relative adjustment of $J_F$ and $J_B$ are available
\cite{jaksch05,takasu03,takamoto03}. 
Here, $E_R = \hbar^2k^2/2m$ is the single photon recoil energy seen by the
$^{40}K$ atoms. 
With the above value of $J_F$, the relevant time scale $\hbar/J_F \simeq 0.3\ \rm s$
is easily resolved. 
Finally, for our subsequent theoretical/numerical treatment, the total
dimension $\cal N$ of the Hilbert space 
scales as  ${\cal N}=L{\cal N}_B$, where ${\cal N}_B=(N+L-1)!/N!(L-1)!$ is the
dimension of the 
bosonic subspace. The exponential increase of $\cal{N}$ with $N_B$ and $L$ is
thus evident. 

In the absence of interactions between the fermionic and bosonic component
($W_{\rm FB}=0$ in
(\ref{4})), the fermions will undergo coherent
Bloch oscillations with a period $T_B=2\pi\hbar/Fd$
\cite{Anderson92}. However, finite
values of $W_{\rm FB}$ will induce a
collision-induced relaxation
of the fermionic dynamics, eventually resulting in a nonvanishing current
across the lattice.
For weak coupling between fermions and bosons,
we can describe the reduced fermionic dynamics in second order perturbation theory in
$W_{\rm FB}$. Further we use the fact that the bosonic
subsystem
is known to exhibit chaotic level
statistics (in the random matrix sense), over a broad range of $W_B$ and $J_B$
\cite{bohigas91,akabepl04}, and
therefore can substitute for a noisy environment
\cite{kolovsky94}, provided its initial state is characterized by equally
distributed populations of the different sublevels (this is, in the
high-temperature limit).
Indeed, the boson-number cross-correlation function
\begin{equation}
\label{7}
R_{l,m}(t,t')=Tr[\hat{n}_l(t)\hat{n}_m(t')] \;,
\end{equation}
\begin{displaymath}
\hat{n}_l(t)=\exp(iH_B t/\hbar)\hat{n}_l\exp(-iH_B t/\hbar) \;,
\end{displaymath}
is very well approximated by the cross correlation generated by a random matrix
Hamiltonian (chosen from the Gaussian Orthogonal Ensemble \cite{bohigas91}), as
illustrated in Fig.~1 for $R_{1,1}(t,0)$ and $R_{1,2}(t,0)$.
\begin{figure}[t]
\center
\includegraphics[width=8 cm]{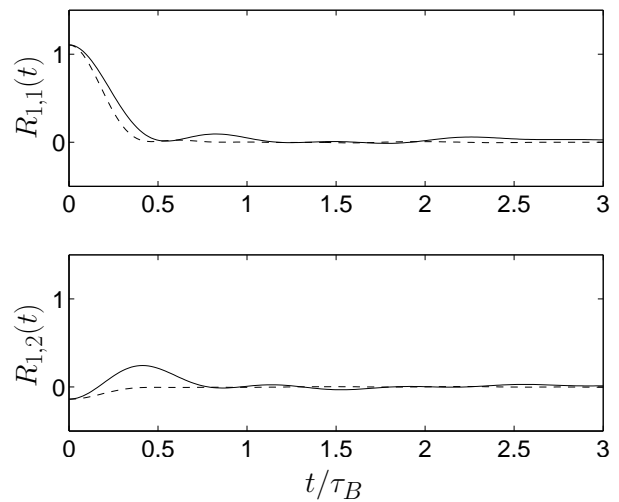}
\caption{Time evolution of the number
correlation functions $R_{1,1}(t,0)$ (top) and
$R_{1,2}(t,0)$ (bottom) of the bosonic bath (\ref{3}), initially prepared in an equally
weigthed superposition of the bosonic eigenstates.
$N=7$, $L=9$.
The ratio between interatomic interaction strength and tunnelling coupling is fixed at
$W_B/J_B=3/7$,
what insures that we are in the parameter range of
Wigner-Dyson spectral statistics, at $\bar{n}\simeq 1$ \cite{akabepl04}.
Time is measured in units of the site-to-site tunnelling period $\tau_B = h/J_B$.
The dashed lines show the bosonic correlation functions which result from
substituting the Bose-Hubbard
Hamiltonian (\ref{3}) by a random matrix of the Gaussian Orthogonal
Ensemble.
}

\label{fig1}
\end{figure}
For $N>2$ and $L$ such that ${\cal N}_B > 100$,
we numerically extract a well defined, finite time scale $\tau\simeq
3\hbar/J_B$ for the decay of these correlations \footnote{For
smaller particle numbers and/or lattice 
sizes, the bath correlations exhibit (partial) revivals on short time scales
which become comparable to the typical time scales we will be considering
here. Hence, in this limit finite size effects would invalidate our subsequent
Markovian treatment.}.

Consequently, if we choose lattice constant $d$ and static field strength
$F$ such that $\tau\ll T_B$ \footnote{The latter implies $Fd \ll 2\pi
J_B/3$, by virtue of the definitions of $\tau$ and $T_B$, and is thus 
compatible 
with our earlier choices $Fd < J_F$ and $J_B=J_F$.},
the correlation function
of the bath  
can be approximated by a $\delta$-function
%****************************************************************
\begin{equation}
\label{8}
R_{l,m}(t,t')=\bar{n}^2\delta_{l,m}\delta\left(\frac{t-t'}{\tau}\right) \;,
\end{equation}
on the typical time scale of the fermionic
dynamics, given by the Bloch period $T_B$.
This justifies a Markov approximation after
tracing out the bosonic degrees of freedom, and
we end up with the master equation for the fermionic density matrix,
\begin{equation}
\label{9}
\frac{\partial \rho^{(F)}_{l,m}}{\partial t}=
-\frac{i}{\hbar}\left[H_F(t),\rho^{(F)}\right]_{l,m}
-\gamma(1-\delta_{l,m})\rho^{(F)}_{l,m} \;,
\end{equation}
where the decay rate is given by 
\begin{equation}
\label{10}
\gamma = \frac{\tau\bar{n}^2 W_{\rm FB}^2}{\hbar^2} \simeq \frac{3\bar{n}^2
W_{\rm FB}^2}{\hbar J_B} \;.
\end{equation}

An analytic solution \cite{kolovsky02} of (\ref{9}) 
leads to the following expression for the mean fermionic
velocity in the lattice:
\begin{equation}
\label{11}
v(t)=v_0e^{-\gamma t}\sin(\omega_B t) \;, \quad
v_0=\frac{J_Fd}{\hbar} \;,
\end{equation}
in perfect agreement with the numerical result
obtained from a direct propagation of the dynamics, with the bosonic bath prepared in an equally
weighted superposition of the eigenstates of the Bose-Hubbard Hamiltonian, garnished by
random phases (tantamount of the high temperature limit), while the fermionic
subsystem is launched in a Bloch wave with 
vanishing quasimomentum. 
As illustrated in Fig.~2, the fermionic Bloch oscillations 
decay irreversibly, with a decay constant $\gamma$ which -- in agreement with
Eq.~(\ref{10}) -- is fully
controlled by the filling factor $\bar{n}$,
by the fermion-boson interaction strength $W_{\rm FB}$, and by the
bosonic tunnelling coupling $J_B$.
\begin{figure}[t]
\center
\includegraphics[width=8 cm]{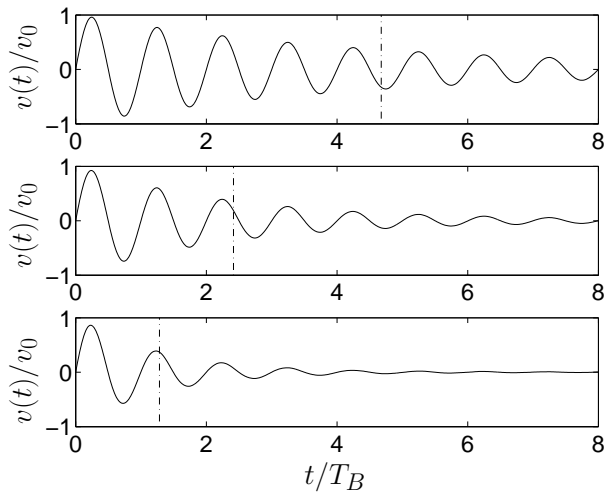}
\caption {Normalized mean velocity of the Fermi atoms, in units of
$v_0=J_Fd/\hbar$, for $Fd=0.57\times J_F$, $J_B=J_F$, for
fermion-boson interaction strengths
$W_{\rm FB}=0.101\times J_F$, $0.143\times J_F$, $0.202\times J_F$ (from top to bottom),
and $N=7$, $L=9$. The typical
time scale of the interaction induced decay of the fermionic Bloch
oscillations fits the time scale $\gamma^{-1}$ 
predicted by eq.~(\ref{10}) (dash-dotted vertical
lines) very well. The same observation holds for fixed $W_{\rm
FB}$ and variable $Fd$ (not shown here).
}
\label{fig2}
\end{figure}

However, Fig.~2 indicates no nonvanishing average drift
of the fermions across the lattice, hence no current is observed. This is 
a consequence of our above derivation: 
while the master equation description provides a quite
satisfactory and quantitative 
picture of the collision induced decoherence of the fermionic
Bloch oscillations, it also implies that there is no back-action of the
fermion-boson coupling on the bath (this is the essence of the
Markov approximation). Consequently, no net energy can be
transfered from the fermionic into the bosonic component (the latter being
initially in a fully thermalized state), as clearly manifest
in the time evolution of the average energy of the bosonic component,
represented by the dashed line in the bottom panel of Fig.~3. 
If, instead, we choose a bosonic initial
condition given by  
a low 
temperature Boltzmann distribution
$P(\epsilon_i)\sim\exp(-\epsilon_i/k_BT)$, with temperature
$k_BT\simeq 2.86\times J_B$ \footnote{It is
possible to show that $|\epsilon_i|\leq NJ_B$, in the limit $W_{\rm
FB}\rightarrow 0$ \protect\cite{akabepl04}.},
over its energy levels $\epsilon_i$, the bath can be heated
by the collisions
with the fermions, and thus extract energy from the fermionic component.  
Such choice of the bosonic initial state leads to
the
relaxation of the fermionic Bloch oscillations, associated with a
(non-Markovian)  energy
increase of the 
bosonic component, see Fig.~3.
\begin{figure}[t]
\center
\includegraphics[width=8 cm]{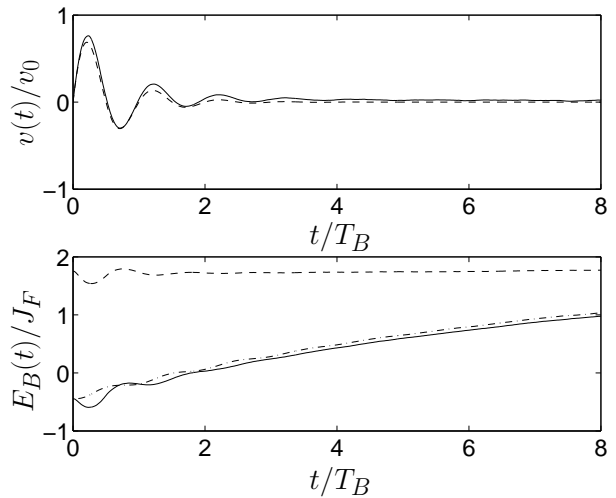}
\caption{Mean velocity $v(t)$  of the fermionic component (top, solid line), 
in units of $v_0=J_Fd/\hbar$, and average energy $E_B(t)$ of
the bosonic component (bottom, solid line), in units of $J_F$, for a low
temperature ($k_B T\simeq 2.86 \times J_B$) initial state of the bosonic bath.
$Fd=0.143\times J_F$, $W_{\rm FB}=0.143\times J_F$, $W_B/J_B = 3/7$, $J_B=J_F$,
$N=7$, $L=9$. The dashed lines indicate the corresponding solutions for a
thermalized initial state of the bath, $k_B T\simeq 150\times J_B$.
For a nonequilibrium initial condition of the bath the finite drift velocity
$\bar{v}(t)$ manifests as a clear, nonvanishing offset of the fermionic mean velocity with respect to
the Markovian result indicated by the dashed line (top). The dash-dotted line in the bottom plot is obtained
by integration of eq.~(\ref{13}).}

\label{fig3}
\end{figure}
Correspondingly, the fermionic mean velocity exhibits 
a nonvanishing drift, on top of the oscillatory behaviour due to the Bloch
dynamics. A finite fermionic current across the lattice is observed in the
upper panel of Fig.~3! Since we will be
dealing with a closed system of finite size (finite particle number and finite
lattice length), the bosonic bath has a finite heat capacity and  
cannot act as a reservoir over arbitrary time
scales. Consequently, the relaxation-induced current ceases as soon as the
bath is 
fully 
thermalized by
draining energy from the fermionic dynamics. 

Fermionic drift 
\begin{equation}
\label{12}
\bar{v}(t)=v(t)-v_0 e^{-\gamma t}\sin(\omega_B t) 
\end{equation}
and bosonic energy gain $\Delta E_B$ are related through the classical
relation $\Delta E_B=F\Delta x=F\bar{v}\Delta t$, i.e.,
\begin{equation}
\label{13}
\frac{\partial E_B}{\partial t}=F\bar{v}(t) \;.
\end{equation}
Integration of (\ref{13}) with $\bar{v}$ as defined in (\ref{12}),
together with $v(t)$ from the exact
solution of the Schr\"odinger equation under
(\ref{1}), leads to a very good fit of the exact time dependence of the bosonic
mean energy, in the lower panel of Fig.~3.
We can thus deduce the current-voltage characteristics for the fermions,
i.e., the dependence of the drift velocity $\bar{v}$ (equivalent to
the current, modulo the carrier density) on the static force (or potential
difference across the lattice) which generates the current:
For sufficiently short
times, eq.~(\ref{13}) can be rewritten as 
$\Delta E_B=F\bar{v}\Delta t$, where $F\bar{v}$ can be extracted from the
exact result for $E_B(t)$ as displayed in Fig.~3 (averaging over the
residual Bloch oscillations during the 
first two Bloch cycles). Dividing
this initial growth rate of $E_B(t)$ by $F$, we obtain the dependence of the 
``current'' $\bar{v}$ on the ``voltage'' 
$F$ in Fig.~4, which exhibits a marked transition
from Ohmic ($\bar{v}\sim F$) behaviour to negative differential conductivity
($\bar{v}\sim 1/F$), as $F$ is increased. 
\begin{figure}[t]
\center
\includegraphics[width=8 cm]{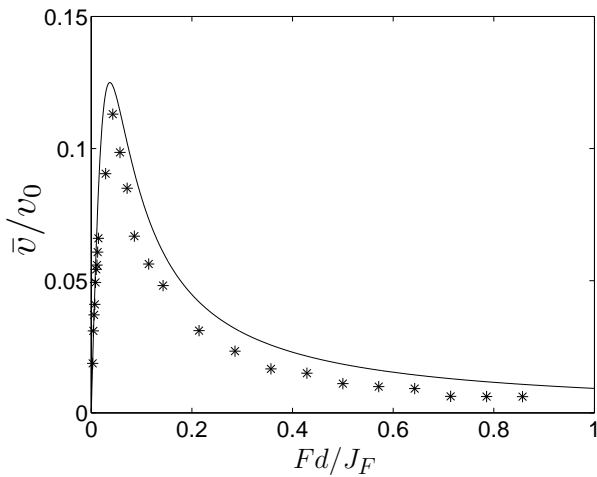}
\caption{Drift velocity $\bar{v}$ (stars), in units of
  $v_0=J_Fd/\hbar$, vs. 
static forcing $Fd$, in units of $J_F$. The Bloch frequency $\omega_B\sim Fd$
increases with increasing $F$, while the effective decay rate $\gamma\simeq 0.037 J_F/\hbar$,
derived from (\ref{10}) for $N=7$, $L=9$, $W_{\rm FB}=0.143\times J_F$,
$W_B/J_B=3/7$, $J_B=J_F$, is constant. Note the clear 
transition from  Ohmic behaviour ($\bar{v}\sim Fd$) to negative differential conductance
($\bar{v}\sim 1/Fd$), at a finite value of $Fd$.
}
\label{fig4}
\end{figure}
Remarkably, this faithfully
reproduces the qualitative behaviour predicted by the 
phenomenalogical Esaki-Tsu relation \cite{esaki70} for a current across
doped superlattices, 
\begin{equation}
\label{16}
\bar{v}=\frac{v_0}{4}\frac{\omega_B/\gamma}{1+(\omega_B/\gamma)^2} \;,
\end{equation}
with quantitative differences, notably for large $F$, due to the finite size
of our atomic sample.
Note that, in contrast to experiments in
semiconductor superlattices, the effective scattering rate $\gamma$ is
perfectly  
controlled and continuously tunable (e.g., via Feshbach resonances
\cite{Ketterle98}) 
in our present quantum optical setting, as a consequence of
eq.~(\ref{10}).

In conclusion, we have shown that a directed current of spinpolarized ultracold fermionic
atoms across a one dimensional optical
lattice under static forcing can be induced by collisional interaction 
with a bosonic admixture.
While
the current increases linearly with the static force in the limit where the
collision induced decay rate (independent of the static forcing!) 
is much larger than the fermionic Bloch frequency (increasing linearly with
the static field strength), it decreases inversely proportional to $F$ in the
opposite limit. This crossover has an intuitive cause: Finite, collision
induced decay rates induce diffusive transport at arbitrarily weak forcing 
(many collision events during one Bloch cycle), while strong forcing
and small collision rates (few collisions during many Bloch cycles)
essentially reestablish the Bloch oscillations, and suppress the directed
current.  

Our theory, based on a microscopic, strictly
Hamiltonian picture of the many-body dynamics, is amenable to 
experimental scrutiny. It also identifies the microscopic origin of first
experimental observations which already pointed in this direction, on the
basis of a purely phenomenological interpretation \cite{ott04}. Finally, our
model allows for bosonic baths composed of an arbitrary number of
particles. Future work will explore the limit of very small baths, with small
Hilbert space dimensions and a reduced density of states. In this regime,
non-Markovian effects due to the granularity of the bath are expected, which
will be in reach for direct experimental probing.

We acknowledge fruitful discussions with Artem Dudarev, and partial financial support by Deutsche
Forschungsgemeinschaft, within the SPP1116 program.

\end{document}